\documentclass[journal]{IEEEtran}
\usepackage{cite}

\ifCLASSINFOpdf
  \usepackage[pdftex]{graphicx}
\else
\fi
\usepackage{amsmath}
\usepackage{mathptmx}
\usepackage{bm}
\usepackage{mathtools}
\usepackage{cancel}
\usepackage{xcolor}
\usepackage{multirow}
\usepackage{makecell}
\definecolor{chi}{HTML}{AD3E2E}
\allowdisplaybreaks

\begin{document}
%
\title{Comparison of Droop-Based Single-Loop Grid-Forming Wind Turbines: High-Frequency Open-Loop Unstable Behavior and Damping}
%
%
%

\author{Meng Chen,~
        Yufei Xi,~
        Lin Cheng,~
        Xiongfei Wang,~
        and Ioannis Lestas
\thanks{This work was funded by UK Research and Innovation (UKRI) under the UKRI Postdoctoral Fellowship Guarantee Grant [EP/Z001889/1]. The authors would like to acknowledge the Department of Engineering, University of Cambridge, for the support of the Fellowship. The contents reflect only the authors' view and not the views of the University or the UKRI.



}
}

\maketitle

\begin{abstract}
The integration of inverter-interfaced generators introduces new instability phenomena into modern power systems. This paper conducts a comparative analysis of two widely used droop-based grid-forming controls, namely droop control and droop-I control, in wind turbines. Although both approaches provide steady-state reactive power-voltage droop characteristics, their impacts on high-frequency (HF) stability differ significantly. Firstly, on open-loop (OL) comparison reveals that droop-I control alters HF pole locations. The application of Routh's Stability Criterion further analytically demonstrates that such pole shifts inevitably lead to OL instability. This HF OL instability is identified as a structural phenomenon in purely inductive grids and cannot be mitigated through control parameter tuning. As a result, droop-I control significantly degrades HF stability, making conventional gain and phase margins insufficient for evaluating robustness against parameter variations. Then, the performance of established active damping (AD) is assessed for both control schemes. The finding indicates that AD designs effective for droop control may fail to suppress HF resonance under droop-I control due to the presence of unstable OL poles. Case studies performed on the IEEE 14-Bus Test System validate the analysis and emphasize the critical role of HF OL instability in determining the overall power system stability.
\end{abstract}


%
\IEEEpeerreviewmaketitle

\section{Introduction}
%
%
%
%
\IEEEPARstart{T}{he} integration of wind turbines (WTs) can introduce new instability phenomena into modern power systems, which is largely governed by the control strategies of their interfacing inverters \cite{Hatziargyriou2021,Blaabjerg2024}. As WTs are increasingly expected to support both frequency and voltage stability in power systems \cite{Saboori2025}, single-loop \cite{Du2020} or multi-loop \cite{Kim2025} grid-forming (GFM) control strategies have been proposed. Regardless of the specific control algorithms, droop characteristics are commonly used in both active power (AP) and reactive power (RAP) loops to enable power-sharing and system support functions \cite{Luo2025}.

In the SL-GFM control, the RAP is typically measured after the filter capacitor, while the generated reference is for the inverter voltage. As a result, applying a direct droop control will lead to errors due to the influence of the filter and any virtual impedance. This issue has been typically addressed by measuring both the RAP and the capacitor voltage to combine the droop characteristics with a proportional-integral (PI) or I controller, referred to in this paper as droop-I control, to achieve robust steady-state droop performance \cite{Chen2024,Luo2025}.

While the stability of low-frequency (LF) power loops in GFM control has been extensively studied \cite{Chen2019,Chen2022,Liu2025}, high-frequency (HF) stability has received comparatively less attention. The HF resonance introduced by the LCL filter is a well-known instability issue associated with conventional grid-following (GFL) control, which has been extensively studied over the past decades \cite{Wang2016,Jayalath2017,Liu2020}. To mitigate such resonance, a variety of active damping (AD) strategies have been proposed, typically using feedback from inverter current, grid current, filter capacitor voltage or current \cite{Leitner2018,Liu2020,Zhang2022, Wang2022}. However, these existing results need further validation in the GFM application.

Although some studies have applied root locus analysis to detailed models that account for HF dynamics, such methods contribute limited insight into the fundamental instability mechanisms from a frequency-domain perspective \cite{Du2020,Park2024,Liu2024a}. Moreover, these studies primarily focus on droop control. A recent frequency-domain analysis of SL-GFM control explicitly links HF instability to the filter capacitor and the resonant frequency \cite{Liu2022}, demonstrating that instability arises from insufficient gain and phase margins in the open-loop (OL) Bode plot determined by Bode's 0dB gain and -180$^\circ$ phase crossover criterion \cite{Franklin2019}. This HF instability can be mitigated through AD strategies, provided the OL frequency response is shaped to guarantee positive stability margins \cite{Liu2024}. However, this conclusion holds primarily for droop control, under the implicit assumption that the system remains minimum phase.

Very recent research has shown that SL-GFM converter using droop-I control may exhibit nonminimum phase behavior, with unstable OL poles emerging near resonant frequencies. This fundamentally alters the HF characteristics \cite{Meng2025}. However, existing analyses have largely relied on quantitative root locus evaluations, without fully explaining the underlying mechanism driving the transition to nonminimum phase behavior or its implications for AD design. In particular, a comparative understanding of how this differs from droop control-based systems remains underexplored.

Therefore, this paper conducts a comparative analysis of these two widely used droop-based SL-GFM controls in wind turbines. The main contributions are summarized as follows:
\begin{enumerate}
    \item Unlike existing studies that rely primarily on quantitative analysis, this paper provides an analytical presentation that the additional voltage feedback in droop-I control fundamentally alters the locations of HF poles in the OL system. This alteration is unique to the droop-I-based RAP control loop and cannot be observed in systems using droop control. Moreover, it cannot be identified through analysis of the decoupled AP loop model, even when the LCL filter dynamics are taken into account.
    \item By applying Routh's Stability Criterion, this paper further confirms, through analytical means, that the HF pole shifts introduced by droop-I control inevitably lead to OL instability. This behavior is identified as the fundamental mechanism underlying HF instability in SL-GFM with droop-I control. Notably, this OL instability emerges as a structural characteristic in purely inductive grids and cannot be mitigated through parameter tuning. Consequently, droop-I control substantially compromises HF stability compared to droop control and makes the stability margins less effective for evaluating robustness under parameter variations. This is because the instability arises from a transition from minimum phase to nonminimum phase with unstable OL poles, rather than from violations of Bode's crossover criterion.
    \item The performance of various established AD strategies is evaluated for both droop and droop-I control schemes. The results indicate that AD designs effective in stabilizing droop-controlled systems may fail to suppress HF resonance when applied to droop-I control. This limitation is primarily due to the presence of unstable OL poles. In the case of droop-I control, a larger damping is required to shift the system dynamics from nonminimum phase back to minimum phase, thereby restoring stability.
\end{enumerate}

The remainder of this paper is organized as follows: Section \ref{Sec_comparison} presents a comparative analysis of the OL behaviors of droop and droop-I control strategies in SL-GFM converters, including an analytical investigation into the mechanism responsible for OL instability at high frequencies. Section \ref{sec_comparison_AD} examines the performance of typical AD designs when applied to both droop and droop-I controls. Section \ref{sec_Case} provides comprehensive time-domain simulation studies using the IEEE 14-Bus Test System, highlighting the differences in system response under droop and droop-I controls. Finally, Section \ref{sec_conclusion} concludes the paper by summarizing the key findings and their implications for SL-GFM converter design.


\section{Comparison of SL-GFM Converter with Droop and Droop-I control Strategies}\label{Sec_comparison}

\subsection{System Description}
\begin{figure*}[!t]
\centering
\includegraphics[width=\textwidth]{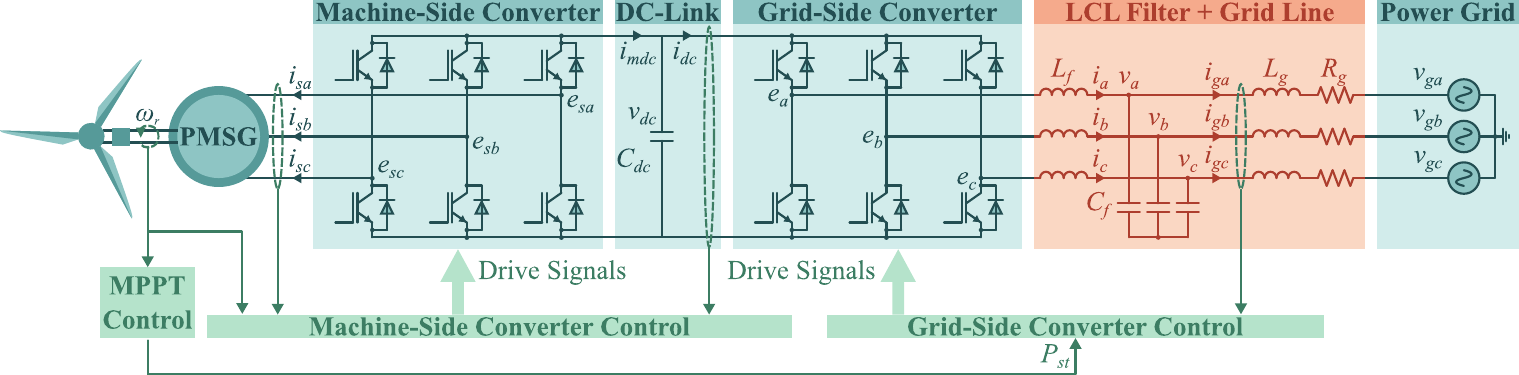}
\caption{Typical configuration of grid-connected WT system.}
\label{fig_Topology}
\end{figure*}

\begin{figure}[!t]
\centering
\includegraphics[width=\columnwidth]{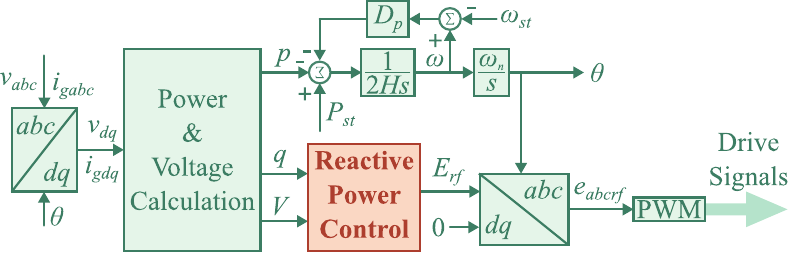}
\caption{Control block diagram of SL-GFM control strategy.}
\label{fig_SLGFM}
\end{figure}

Fig. \ref{fig_Topology} illustrates a typical configuration of a grid-connected WT system. When the grid-side converter is operated under the SL-GFM control strategy, as shown in Fig. \ref{fig_SLGFM}, the virtual swing equation is typically used for the AP control, expressed in p.u. value as
\begin{align}
    2H\frac{d\omega}{dt} = P_{st} - p - D_p(\omega - \omega_{st})
\end{align}
where $H$ is the inertia constant, $\omega$ is the angular frequency, $p$ is the output AP, and $D_p$ is the droop coefficient of the AP control loop. The subscript "$st$" denotes the set-point value of the corresponding variables.

\begin{figure}[!t]
\centering
\includegraphics[width=\columnwidth]{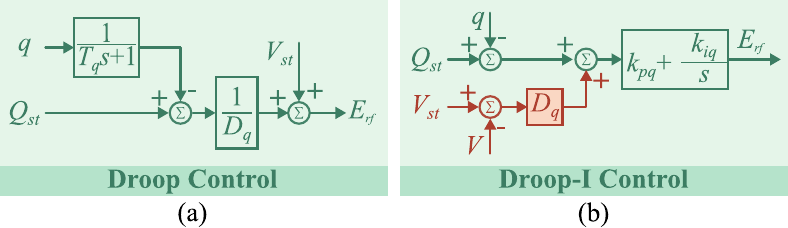}
\caption{Control block diagram of RAP control. (a) Droop control. (b) Droop-I control.}
\label{fig_RAP}
\end{figure}

Two droop-based structures are widely used for implementing the RAP control loop, namely droop control and droop-I control, as shown in Fig. \ref{fig_RAP}. The generated inverter voltage magnitude reference $E_{rf}$ from the two strategies can be expressed, respectively, as
\begin{align}
    &E_{rf} = V_{st} + \frac{1}{D_q}\left(Q_{st} - \frac{1}{T_qs + 1}q\right)\\
    &E_{rf} = \left(k_{pq} + \frac{k_{iq}}{s}\right)[D_q(V_{st} - V) + Q_{st} - q]
\end{align}
where $V$ is the capacitor voltage magnitude, $D_q$ is the droop coefficient of the RAP control loop, $T_q$ is the time constant of the RAP filter in droop control, $q$ is the output RAP, $k_{pq}$ and $k_{iq}$ are the proportional and integral gains of the PI controller in droop-I control. In many studies, $k_{pq}=0$ is selected, which does not change the desired droop characteristics. A notable advantage of the droop-I control over the droop control is its ability to maintain an accurate $q$-$V$ droop relationship thanks to the PI controller, regardless of the influence of the filter or any virtual impedance.

\subsection{Comparatively Analysis on Open-Loop Models}

Although both the droop control and droop-I control yield the droop behavior, this section shows that their HF stability characteristics differ substantially.

\begin{figure}[!t]
\centering
\includegraphics[width=\columnwidth]{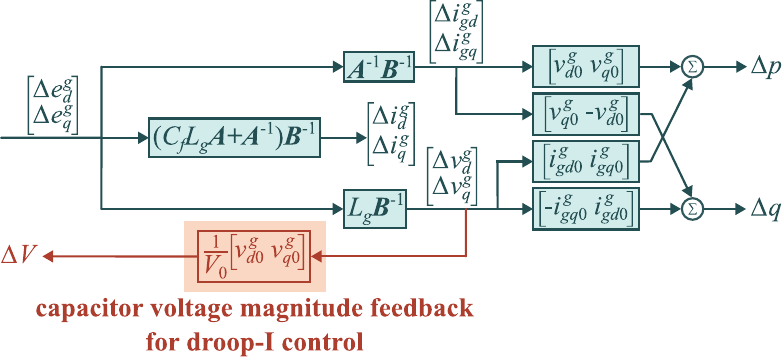}
\caption{Small-signal block diagram of filter and grid line.}
\label{fig_LCLSmallSignal}
\end{figure}

When neglecting the resistors, the filter and grid line dynamics governing the HF behavior can be described as
\begin{align}
    & \frac{L_f}{\omega_n}\dot i^g_d = e^g_d - v^g_d + \omega_g L_fi^g_q\\
    & \frac{L_f}{\omega_n}\dot i^g_q = e^g_q - v^g_q - \omega_g L_fi^g_d\\
    & \frac{C_f}{\omega_n}\dot v^g_d = i^g_d - i^g_{gd} + \omega_g C_fv^g_q\\
    & \frac{C_f}{\omega_n}\dot v^g_q = i^g_q - i^g_{gq} - \omega_g C_fv^g_d\\
    & \frac{L_g}{\omega_n}\dot i^g_{gd} = v^g_d - V_g + \omega_g L_gi^g_{gq}\\
    & \frac{L_g}{\omega_n}\dot i^g_{gq} = v^g_q - \omega_g L_gi^g_{gd}\\
    & p = v^g_di^g_{gd} + v^g_qi^g_{gq}\\
    & q = v^g_qi^g_{gd} - v^g_di^g_{gq}\\
    & V = \sqrt{(v^g_d)^2 + (v^g_q)^2}
\end{align}
where $L_f$ and $L_g$ are the inverter-side and grid-side inductances, $C_f$ is the filter capacitance, $i$ and $i_g$ represent the inverter and grid currents, $e$ and $v$ are the inverter and capacitor voltages, $\omega_n$ is the nominal angular frequency, $V_g$ is the grid voltage magnitude, and $\omega_g$ is the grid angular frequency. Subscripts "$d$" and "$q$" indicate the components in the $dq$ frame, and the superscript "$g$" denotes the grid synchronous reference frame. The corresponding small-signal block diagram is shown in Fig. \ref{fig_LCLSmallSignal}, based on which the outputs can be calculated by
\begin{align}
\label{eq_Deltap}
    & \Delta p = \left(
      \begin{bmatrix}
          v_{d0}^g & v_{q0}^g
      \end{bmatrix}
      \bm{A}^{-1} + L_g
      \begin{bmatrix}
          i_{gd0}^g & i_{gq0}^g
      \end{bmatrix}
      \right)\bm{B}^{-1}
      \begin{bmatrix}
          \Delta e_d^g\\
          \Delta e_q^g
      \end{bmatrix}\\
    & \Delta q = \left(
      \begin{bmatrix}
          v_{q0}^g & -v_{d0}^g
      \end{bmatrix}
      \bm{A}^{-1} + L_g
      \begin{bmatrix}
          -i_{gq0}^g & i_{gd0}^g
      \end{bmatrix}
      \right)\bm{B}^{-1}
      \begin{bmatrix}
          \Delta e_d^g\\
          \Delta e_q^g
      \end{bmatrix}\\
\label{eq_DeltaV}
    & \Delta V = \frac{L_g}{V_0}
      \begin{bmatrix}
          v_{d0}^g & v_{q0}^g
      \end{bmatrix}
      \bm{B}^{-1}
      \begin{bmatrix}
          \Delta e_d^g\\
          \Delta e_q^g
      \end{bmatrix}.
\end{align}
with the subscript "0" denoting steady-state values. The matrices $\bm A$ and $\bm B$ are defined as
\begin{align}
   & \bm{A} \coloneqq
     \frac{1}{\omega_n}\begin{bmatrix}
          s & -\omega_{Lg}\\
          \omega_{Lg} & s
      \end{bmatrix}\\
   & \bm{B} \coloneqq L_fC_fL_g\bm{A}^2 + (L_f + L_g)\bm{I}_2
\end{align}
where $\omega_{L_g}$ is the synchronous frequency, defined as
\begin{align}
    \omega_{L_g} \coloneqq \omega_n\omega_g.
\end{align}

As illustrated in Fig. \ref{fig_LCLSmallSignal}, droop-I control incorporates an additional feedback branch using the capacitor voltage magnitude. The analysis that follows demonstrates that this extra feedback path exerts a significant influence on the system's HF behavior. 

In (\ref{eq_Deltap})-(\ref{eq_DeltaV}), the inverter voltages can be expressed in terms of the SL-GFM control outputs as
\begin{align}
    \begin{bmatrix}
        e^g_d & e^g_q
    \end{bmatrix}^T = E_{rf}
    \begin{bmatrix}
        \cos\delta & \sin\delta
    \end{bmatrix}^T
\end{align}
where the dynamic of the angle difference $\delta$ is described by
\begin{align}
    \dot\delta = \omega_n(\omega-\omega_g).
\end{align}

\begin{figure}[!t]
\centering
\includegraphics[width=\columnwidth]{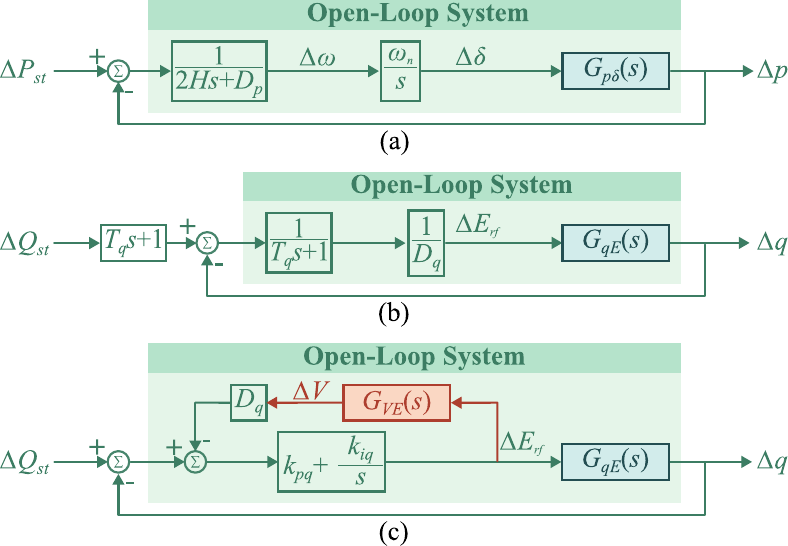}
\caption{Decoupled small-signal block diagrams. (a) AP loop. (b) RAP loop under droop control. (c) RAP loop under droop-I control.}
\label{fig_DecoupledSmallSignal}
\end{figure}

Under the conventional assumption of an inductive grid, the AP and RAP loops can be considered decoupled. The resulting block diagrams of the power loops are shown as Fig. \ref{fig_DecoupledSmallSignal}. In these diagrams, $G_{p\delta}(s)$, $G_{qE}(s)$, and $G_{VE}(s)$ denote the transfer functions between their respective inputs and outputs, defined as
\begin{align}
    & G_{p\delta}(s) \coloneqq \frac{N_{p\delta}(s)}{\rm{det}(\bm{A})\rm{det}(\bm{B})}\\
    & G_{qE}(s) \coloneqq \frac{N_{qE}(s)}{\rm{det}(\bm{A})\rm{det}(\bm{B})}\\
    & G_{VE}(s) \coloneqq \frac{N_{VE}(s)}{\rm{det}(\bm{B})}
\end{align}
where det($\cdot$) denotes the matrix determinant, and $N_{p\delta}(s)$, $N_{qE}(s)$, and $N_{VE}(s)$ are the corresponding numerator polynomials, defined as
\begin{align}
     N_{p\delta}(s) \coloneqq &\left(
      \begin{bmatrix}
          v_{d0}^g & v_{q0}^g
      \end{bmatrix}
      \bm{A}^T + L_g\rm{det}(\bm{A})
      \begin{bmatrix}
          i_{gd0}^g & i_{gq0}^g
      \end{bmatrix}
      \right)\bm{B}^T
      \begin{bmatrix}
        -e_{q0}^g\\
        e_{d0}^g
      \end{bmatrix}\\
      N_{qE}(s) \coloneqq &\left(
      \begin{bmatrix}
          v_{q0}^g & -v_{d0}^g
      \end{bmatrix}
      \bm{A}^{T} + L_g\rm{det}(\bm{A})
      \begin{bmatrix}
          -i_{gq0}^g & i_{gd0}^g
      \end{bmatrix}
      \right)\notag\\
     &\bm{B}^{T}
      \begin{bmatrix}
        \cos\delta_0\\
        \sin\delta_0
      \end{bmatrix}\\
     N_{VE}(s) \coloneqq &\frac{L_fC_fL_g^2}{V_0\omega_n^2}[(s^2 + \omega_1\omega_2)v_{d0} - 2\omega_{L_g}v_{q0}s].
\end{align}

As shown in Fig. \ref{fig_DecoupledSmallSignal}(b), the OL model of the RAP loop under droop control can be defined as
\begin{align}
\label{eq_G_RA_droop}
    G_{RA\_droop}(s) \coloneqq \frac{N_{qE}(s)}{D_q(T_qs + 1)\rm{det}(\bm{A})\rm{det}(\bm{B})}.
\end{align}
The determinants of $\bm A$ and $\bm B$ can be written as
\begin{align}
    & {\rm{det}}(\bm{A}) \coloneqq \frac{s^2}{\omega_n^2} + \omega_g^2\\
    & {\rm{det}}(\bm{B}) \coloneqq \frac{L_g^2}{\omega_{LC}^4}(s^2 + \omega_1^2)(s^2 + \omega_2^2)
\end{align}
where $\omega_{LC}$ denotes the LC resonant frequency, $\omega_1$ and $\omega_2$ denote the LCL resonant frequencies in $dq$ frame, defined as
\begin{align}
    & \omega_{LC} \coloneqq \omega_n\sqrt{\frac{1}{L_fC_f}}\\
    & \omega_1 \coloneqq \omega_{LCL} - \omega_{L_g}\\
    & \omega_2 \coloneqq \omega_{LCL} + \omega_{L_g}.
\end{align}
Here, $\omega_{LCL}$ is the LCL resonant frequency, defined as
\begin{align}
\label{eq_wLCL}
    \omega_{LCL} \coloneqq \omega_n\sqrt{\frac{L_f + L_g}{L_fC_fL_g}}.
\end{align}

It follows from (\ref{eq_G_RA_droop})-(\ref{eq_wLCL}) that the RAP loop is entirely decoupled from the resonance. Specifically, the OL poles corresponding to resonance are precisely at the system's natural resonant frequencies $\omega_1$, $\omega_2$, and $\omega_{L_g}$ by setting det($\bm{A}$)det($\bm{B})=0$. These poles lie on the imaginary axis, indicating marginal stability in the ideal pure inductive case. In practical systems, parasitic resistances shift them into the left-half plane (LHP), ensuring the power loops remain minimum phase. Under these conditions, Bode's crossover criterion provides a reliable means for performance evaluation. These results are consistent with previously reported results for SL-GFM converters under droop control \cite{Liu2022,Liu2024}.

However, under droop-I control, the inclusion of the voltage magnitude $\Delta V$ in the RAP loop feedback fundamentally changes the system behavior. In particular, nonminimum phase characteristics can arise due to appearance of unstable OL poles. As shown in Fig. \ref{fig_DecoupledSmallSignal}(c), the OL model of the RAP loop under droop-I control can be defined as
\begin{align}
\label{eq_G_RA_droopI}
    G_{RA\_droopI}(s) \coloneqq& \frac{k_{pq}s + k_{iq}}{s + (k_{pq}s + k_{iq})D_qG_{VE}(s)}G_{qE}(s)\notag\\
     =&\frac{(k_{pq}s + k_{iq})\color{chi}\bcancel{\rm{det}(\bm{B})}}{s{\rm{det}}(\bm{B}) + (k_{pq}s + k_{iq})D_qN_{VE}(s)}\frac{N_{qE}(s)}{\rm{det}(\bm{A})\color{chi}\bcancel{\rm{det}(\bm{B})}}\notag\\
     =& \frac{(k_{pq}s + k_{iq})N_{qE}(s)}{\underbrace{[s{\rm{det}}(\bm{B}) + (k_{pq}s + k_{iq})D_qN_{VE}(s)]}_{\text{new open-loop poles}}{\rm{det}}(\bm{A})}.
\end{align}
This formulation shows that while the OL poles associated with the synchronous frequency given by det($\bm{A}$) remain unchanged, the original OL HF resonance poles from det($\bm{B}$) are canceled due to the feedback interaction. In their place, new OL poles arise from the characteristic equation in the denominator, which determine the OL HF behavior. Meanwhile, the RAP loop is no longer decoupled from the OL HF resonance dynamics. It is worth noting that this OL HF pole-shifting effect is specific to droop-I-based RAP control loop and cannot be observed in droop control, nor in the AP loop of Fig. \ref{fig_DecoupledSmallSignal}(a), even when HF dynamics are explicitly considered.

\subsection{Open-Loop Unstable Behavior in Droop-I Control}

This section presents a detailed analysis on the new OL characteristic equation arising under droop-I control, clarifying the mechanism responsible for the emergence of the OL unstable poles. The resulting shift from minimum phase to nonminimum phase behavior leads to a significant deterioration in HF stability.

As shown in (\ref{eq_G_RA_droopI}), the new OL poles under droop-I control are determined by the characteristic equation
\begin{align}
    s{\rm{det}}(\bm{B}) + (k_{pq}s + k_{iq})D_qN_{VE}(s) = 0
\end{align}
which can be expanded into the fifth-order polynomial form
\begin{align}
\label{eq_CharEq}
    s^5 + a_3s^3 + a_2s^2 + a_1s + a_0 = 0
\end{align}
with coefficients defined as
\begin{align}
    & a_3 \coloneqq \omega_1^2 + \omega_2^2 + \frac{D_qk_{pq}\omega^2_{LC}v_{d0}}{V_0}\\
    & a_2 \coloneqq \frac{D_q\omega_{LC}^2}{V_0}(k_{iq}v_{d0} - 2k_{pq}\omega_{L_g}v_{q0})\\
    & a_1 \coloneqq \omega_1^2\omega_2^2 + \frac{D_q\omega_{LC}^2}{V_0}(\omega_1\omega_2k_{pq}v_{d0}-2k_{iq}\omega_{L_g}v_{q0})\\
    & a_0 \coloneqq \frac{v_{d0}k_{iq}D_q\omega_{LC}^2\omega_1\omega_2}{V_0}.
\end{align}

Note that (\ref{eq_CharEq}) has five roots: two pairs associated with the HF dynamics and a single root corresponding to the LF RAP control. To analyze the OL stability, Routh's Stability Criterion is applied through the following Routh array
\begin{align}
\begin{matrix}
    s^5: & 1 & a_3 & a_1\\
    s^4: & \epsilon & a_2 & a_0\\
    s^3: & a_3 - a_2/\epsilon & a_1 - a_0/\epsilon\\
    s^2: & a_2 - (a_1\epsilon^2 - a_0\epsilon)/(a_3\epsilon - a_2) & a_0\\
    s: &a_1 + \frac{(a_0a_3^2 - a_0a_1)\epsilon + a_0^2 - a_0a_2a_3}{a_1\epsilon^2 - (a_0 + a_2a_3)\epsilon + a_2^2}\\
    s^0:& a_0
\end{matrix}\notag   
\end{align}
Here, a small positive parameter $\epsilon>0$ is introduced in place of the zero coefficient in the $s^4$ row. This perturbation serves a dual purpose: mathematically, it ensures the Routh array is properly defined; physically, it reflects the influence of practical parasitic resistances present in the system.

Evaluating the first few entries in the first column of the Routh array gives
\begin{align}
    &\lim_{\epsilon\rightarrow0^+}\left(a_3 - \frac{a_2}{\epsilon}\right) = -\infty < 0\\
    &\lim_{\epsilon\rightarrow0^+}\left(a_2 - \frac{a_1\epsilon^2 - a_0\epsilon}{a_3\epsilon - a_2}\right) = a_2 > 0.
\end{align}
These results indicate that, assuming an ideally inductive grid, the system inevitably has at least two poles in the right-half plane (RHP), regardless of the signs of other terms in the Routh array. Consequently, the OL HF instability always exists. Notably, this implies a fundamental structural OL instability introduced by the feedback loop involving $\Delta V$. Importantly, such instability cannot be eliminated by adjusting controller and system parameters in an ideally inductive grid.

In practical systems, where the grid is not perfectly inductive and $\epsilon$ is a small finite positive number, the term $(a_3-a_2/\epsilon)$ can still become negative. Notably, as the RAP loop parameter $k_{iq}$ increases, this term tends to decrease, potentially driving the OL poles into instability. This reveals a critical coupling between the HF resonance and the RAP loop, where increasing the RAP loop bandwidth can make the system endure HF instability. Importantly, this type of closed-loop instability is not a result of violating Bode's crossover criterion. Instead, it originates from OL unstable poles. Moreover, the Routh stability result depends not only on the LCL resonant frequency $\omega_{LCL}$ but also on the LC resonance frequency $\omega_{LC}$. This means that the HF behavior under droop-I control is tied to specific parameter values of the system, rather than being simply characterized by $\omega_{LCL}$. Therefore, the conventional study based only on $\omega_{LCL}$ in the context of droop control cannot be directly applied to droop-I control. These analytical insights support and explain the quantitative results reported in \cite{Meng2025}.

The HF stability of the SL-GFM converter under both droop and droop-I control is further compared to validate the preceding analysis. The system parameters are listed in Table \ref{tab_Parameter}, and three comparison cases are carried out.

\begin{table}[!t]
    \centering
    \caption{Parameters of SL-GFM PMSG-WT System}
    \begin{tabular}{clc}
        \hline\hline
        Parameters & \multicolumn{1}{c}{Description} & Values \\ \hline
        $S_n$ & Nominal power & 5 MW (1 p.u.)\\ 
        $V_n$ & Nominal line-to-line RMS voltage & 690 V (1 p.u.)\\ 
        $f_n$ & Nominal frequency & 50 Hz (1 p.u.)\\ 
        $C_f$ & Filter capacitor & 1.6 mF\\
        $L_f$ & Inverter-side filter inductor & 32 $\mu$H\\
        $L_g$ & Grid-side filter inductor + line & 60 $\mu$H\\
        $X_g/R_g$ & $X/R$ ratio of the line & 8\\
        $H$ & Inertia constant & 0.5 s\\
        $D_p$ & Damping of active power control & 50\\
        $D_q$ & Droop coefficient of $q-V$ control & 10\\
        $\omega_{st}$ & Set-point of angular frequency & 1 p.u.\\
        $Q_{st}$ & Set-point of reactive power & 0 p.u.\\
        $V_{st}$ & Set-point of voltage magnitude & 1 p.u.\\
        $k_{pq}$ & Proportional gain of droop-I control & 0\\
        $k_{iq}$ & Integral gain of droop-I control & 4\\
        $T_q$ & Time constant of $q-V$ droop control &0.051\\
        \hline
    \end{tabular}
    \label{tab_Parameter}
\end{table}

\begin{figure}[!t]
\centering
\includegraphics[width=\columnwidth]{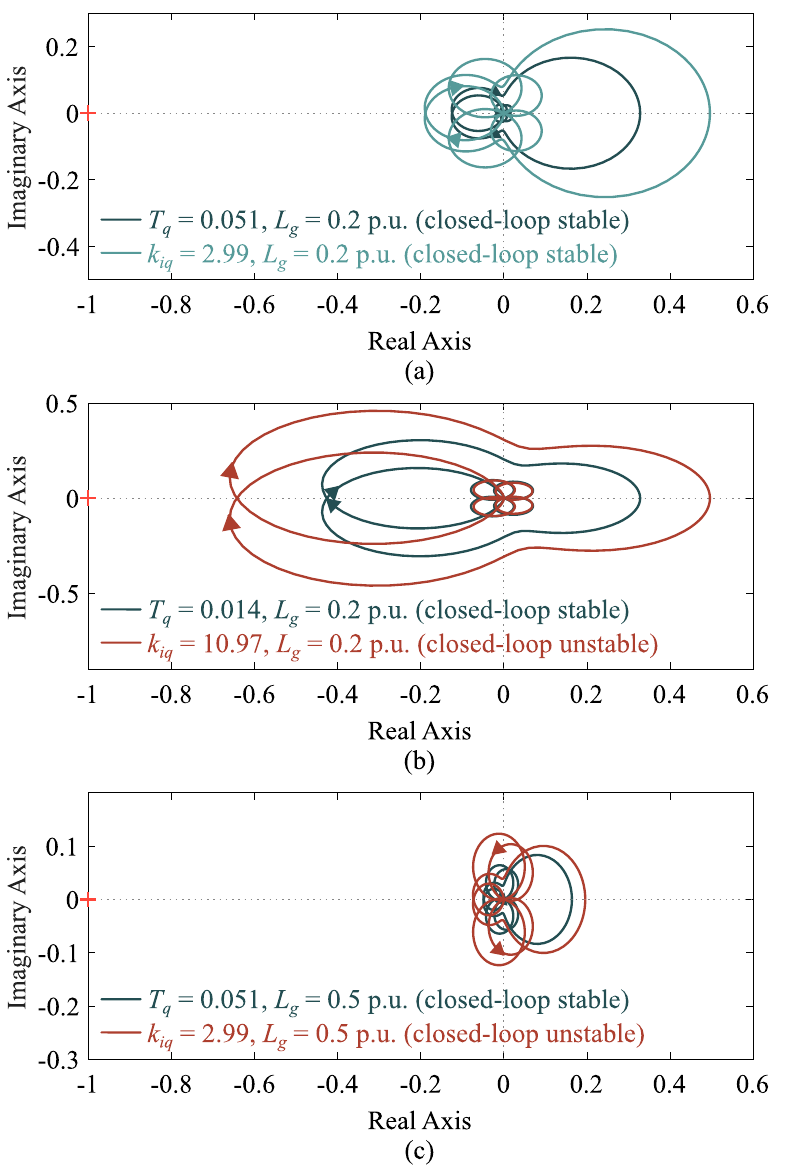}
\caption{Comparison of Nyquist plots of RAP OL systems under droop control and droop-I control.}
\label{fig_NyquistPlot}
\end{figure}

In the first case, $L_g=0.2$ p.u. For a fair comparison, $T_q$ and $k_{iq}$ are set to be 0.051 and 2.99, respectively, ensuring that the OL pole associated with the RAP mode is located at the same value, -19.6, for both droop and droop-I control. The calculated OL poles are presented in Table \ref{tab_OLP}. It can be observed that, under these settings, both control strategies yield minimum phase systems. Their Nyquist plots, shown in Fig. \ref{fig_NyquistPlot}(a), confirm that the systems are stable in this case, as neither plot encircles the critical point (-1, 0).

\begin{table}[!t]
    \centering
    \caption{Comparison of Open-Loop Poles Corresponding to RAP and High-Frequency Resonant Modes under Droop and Droop-I Control Strategies}
    \begin{tabular}{cc||cc}
        \hline\hline
        \multicolumn{2}{c||}{Droop Control} & \multicolumn{2}{c}{Droop-I Control} \\ \hline
        Values & Open-loop poles& Values & Open-loop poles\\\hline
        \multirow{3}{*}{\makecell{$T_q = 0.051$\\$L_g = 0.2$ p.u.}} & -19.6 & \multirow{3}{*}{\makecell{$k_{iq} = 2.99$\\$L_g = 0.2$ p.u.}}& -19.6\\ 
         & $-6.8\pm j5786.6$ & &$-2.2\pm j5786.3$\\ 
         & $-6.8\pm j5158.3$ & &$-1.7\pm j5158.6$\\ \hline
         \multirow{3}{*}{\makecell{$T_q = 0.014$\\$L_g = 0.2$ p.u.}}& -71.4 & \multirow{3}{*}{\makecell{$k_{iq} = 10.97$\\$L_g = 0.2$ p.u.}} & -71.4\\
         & $-6.8\pm j5786.6$ & &\color{chi}{$10.1\pm j5785.3$}\\
         & $-6.8\pm j5158.3$ & &\color{chi}{$12.1\pm j5159.9$}\\\hline
         \multirow{3}{*}{\makecell{$T_q = 0.051$\\$L_g = 0.5$ p.u.}}& -19.6 & \multirow{3}{*}{\makecell{$k_{iq} = 2.99$\\$L_g = 0.5$ p.u.}} & -24.8\\
         & $-3.4\pm j5173.8$ & &\color{chi}{$2.4\pm j5173.4$}\\
         & $-3.4\pm j4545.5$ & &\color{chi}{$3.2\pm j4545.9$}\\
        \hline
    \end{tabular}
    \label{tab_OLP}
\end{table}

In the second case, a larger RAP control bandwidth is considered. Specifically, $T_q$ and $k_{iq}$ are set to 0.014 and 10.97, respectively, shifting the OL RAP mode to -71.4. As expected, Table \ref{tab_OLP} shows that the system under droop control remains minimum phase. However, for droop-I control, the increased RAP bandwidth pushes the OL HF modes into the RHP, causing the system to transition to a nonminimum phase condition. Notably, Fig. \ref{fig_NyquistPlot}(b) shows that both Nyquist plots still do not encircle (-1, 0). This indicates that, while the system under droop control remains stable, the system under droop-I control experiences HF instability.

In the third case, $T_q$ and $k_{iq}$ are kept at their original values of 0.051 and 2.99, respectively. The grid inductance $L_g$ is increased to 0.5 p.u. while maintaining the same $X_g/R_g$ ratio, representing a weaker grid. Likewise, the system under droop control remains minimum phase. In contrast, the system under droop-I control becomes OL unstable under such a weaker grid. As shown in Fig. \ref{fig_NyquistPlot}(c), increasing $L_g$ drives the Nyquist plots further away from (-1, 0). Nevertheless, the system under droop-I control actually becomes HF unstable due to the emergence of RHP OL poles.

The results from the three cases clearly confirm the significantly different HF behaviors of droop and droop-I control strategies. Compared to droop control, droop-I control significantly deteriorates HF stability and exhibits large sensitivity to parameter variations. In such cases, conventional stability margins provide limited insight into robustness, as the instability originates from a minimum phase to nonminimum phase transition caused by unstable OL poles, rather than from violations of Bode's crossover criterion.

\section{Comparison Considering Various AD Strategies}\label{sec_comparison_AD}

\begin{figure}[!t]
\centering
\includegraphics[width=\columnwidth]{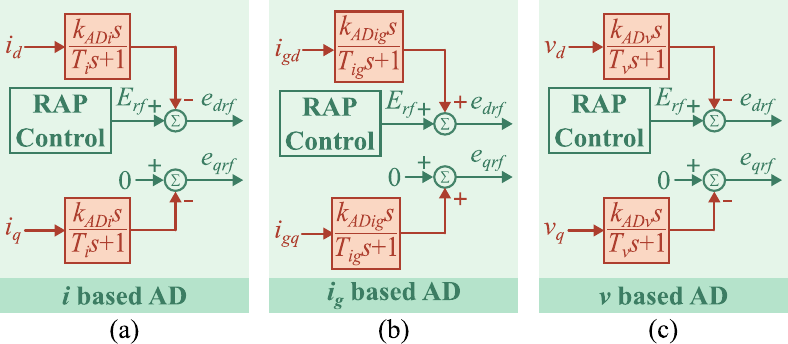}
\caption{Block diagram of AD strategies based on (a) inverter current, (b) grid current, and (c) capacitor voltage.}
\label{fig_AD}
\end{figure}

\begin{figure}[!t]
\centering
\includegraphics[width=\columnwidth]{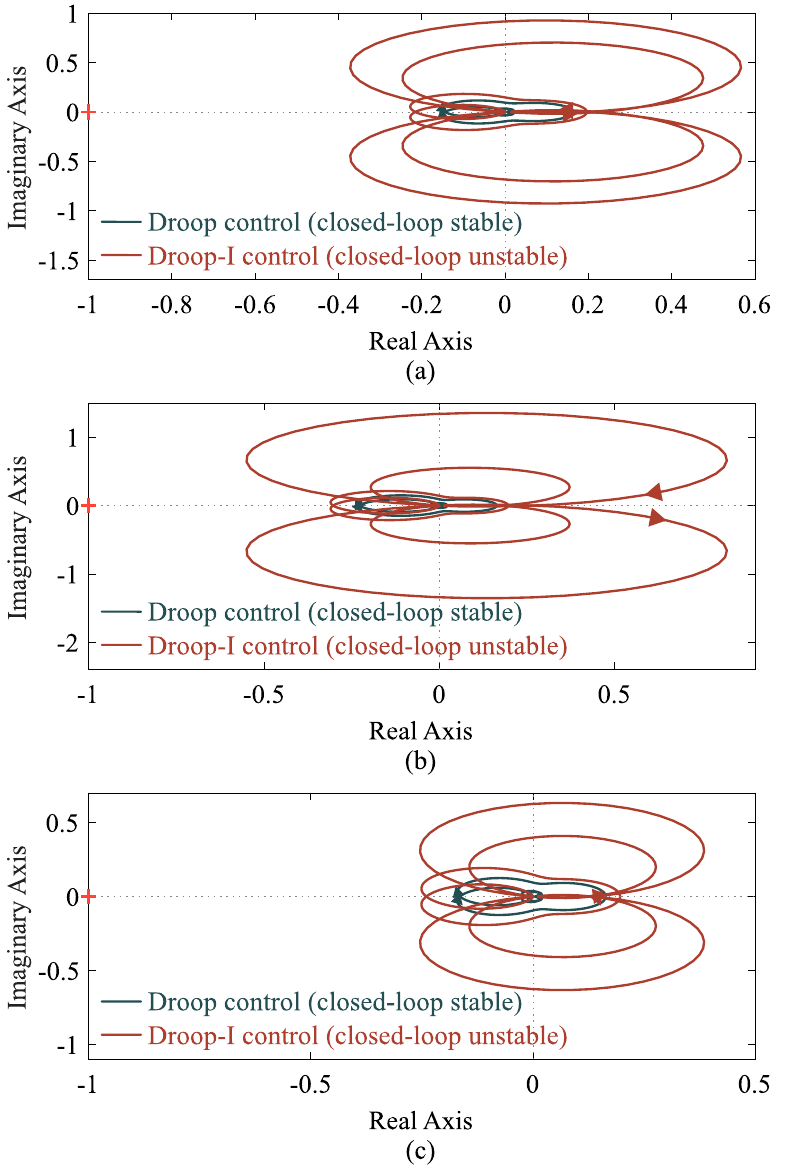}
\caption{Comparison of Nyquist plots under droop control and droop-I control with different AD designs. (a) Inverter current-based AD. (b) Grid current-based AD. (c) Voltage-based AD.}
\label{fig_NyquistAD}
\end{figure}

This section compares the performance of different AD strategies under droop and droop-I control. In SL-GFM control, both the capacitor voltage and the grid current can be used for constructing the AD, thereby avoiding the need for additional sensors \cite{Liu2020,Meng2025}. In addition, since inverter current is explicitly used for SL-GFM AD in \cite{Liu2024}, this approach is also evaluated in this paper. The block diagram of the three AD strategies is presented in Fig. \ref{fig_AD}. The system parameters are the same as those listed in Table \ref{tab_Parameter} except that $T_q = 0.014$, $k_{iq} = 10.97$, and $L_g = 0.5$ p.u. are used, representing the worst case with large RAP control bandwidth and weak grid. The calculated OL poles under droop and droop-I control with different AD designs are summarized in Table \ref{tab_AD}, and the corresponding Nyquist plots are shown in Fig. \ref{fig_NyquistAD}.

\begin{table}[!t]
    \centering
    \caption{Comparison of Open-Loop Poles Corresponding to RAP and High-Frequency Resonant Modes}
    \begin{tabular}{c||c}
        \hline\hline
        \multicolumn{2}{c}{Inverter current-based AD} \\ \hline
        Droop control & Droop-I control\\\hline
        -71.4 & -90.5\\ 
        $-22.5\pm j5172.7$ & $-1.2\pm j5171$\\ 
        $-22.5\pm j4544.3$ & \color{chi}{$1.7\pm j4546.6$}\\ \hline\hline
        \multicolumn{2}{c}{Grid current-based AD} \\ \hline
        Droop control & Droop-I control\\\hline
         -71.4 & -90.9\\
        $-22.1\pm j5171.7$ &$-0.8\pm j5169.8$\\
        $-22.1\pm j4543.1$ & \color{chi}{$2.2\pm j4545.3$}\\\hline\hline
        \multicolumn{2}{c}{Voltage-based AD} \\\hline
        Droop control & Droop-I control\\\hline
        -71&-91\\
        $-23\pm j5182$ &$-1.7\pm j5180$\\
        $-21\pm j4552$ &\color{chi}{$3\pm j4554$}\\
        \hline
    \end{tabular}
    \label{tab_AD}
\end{table}

\subsection{Inverter Current-Based AD Strategy}

As shown in Fig. \ref{fig_AD}(a), in the inverter current-based AD strategy, the inverter currents $i_{dq}$ are fed back to the inverter voltage reference generated by the SL-GFM control through a gain $k_{ADi}$. To eliminate steady-state offsets, a high-pass filter is incorporated into the feedback path \cite{Liu2024}, where $T_i$ denotes the time constant of the filter. A design example of the AD controller, denoted as $G_{ADi}(s)$, used in this paper is defined as:
\begin{align}
\label{eq_GADi}
    G_{ADi}(s) = \frac{5.5\times10^{-5}s}{1/(90\pi)s+1}
\end{align}

Table \ref{tab_AD} shows that the system under droop control with the designed $G_{ADi}(s)$ has HF OL poles located far away from the imaginary axis. Meanwhile, Fig. \ref{fig_NyquistAD}(a) indicates that the Nyquist plot does not encircle the critical point (-1, 0), implying that the system is quite stable at high frequencies. In contrast, under droop-I control with the same $G_{ADi}(s)$, the system remains nonminimum phase with two poles in the RHP. Therefore, although its Nyquist plot also does not encircle (-1, 0), the system is expected to suffer from HF instability. This demonstrates that an inverter current-based AD design which effectively guarantees HF stability of the system under droop control may not be sufficient for the system under droop-I control, if it cannot safely eliminate unstable OL poles.

\subsection{Grid Current-Based AD Strategy}

As shown in Fig. \ref{fig_AD}(b), in the grid current-based AD strategy, the AD controller, denoted as $G_{ADi_g}(s)$, is implemented as a negative high-pass filter acting on the grid currents $i_{gdq}$ to equivalently modifying the circuit characteristics \cite{Wang2016}, where $k_{ADi_g}$ and $T_{i_g}$ are the gain and time constant of the filter, respectively. A design example of the AD controller used in this paper is defined as
\begin{align}
\label{eq_GADig}
    G_{ADi_g}(s) = \frac{1.3\times10^{-4}s}{1/(180\pi)s + 1}
\end{align}

Table III shows that the designed $G_{ADi_g}(s)$ places the OL poles of the system under droop control at positions in the LHP similar to those achieved with $G_{ADi}(s)$. Likewise, the Nyquist plot in Fig. \ref{fig_NyquistAD}(b) indicates large stability margins and stable operation. In contrast, applying the same $G_{ADi_g}(s)$ under droop-I control leaves one pair of HF OL poles in the RHP, with a positive real part of 2.2. This nonminimum phase behavior causes the system to experience HF instability, despite its Nyquist plot not encircling (-1, 0), as shown in Fig. \ref{fig_NyquistAD}(b). These results demonstrate that, similar to the inverter current-based case, a grid current-based AD design that ensures HF stability of the system under droop control may still be insufficient to eliminate unstable OL poles and suppress HF oscillations in a system under droop-I control.

\subsection{Capacitor Voltage-Based AD Strategy}

In the capacitor voltage-based AD strategy, the capacitor voltage $v_{dq}$ are fed back to the inverter voltage reference generated by the SL-GFM control through a pure derivative term $k_{ADv}s$ to equivalently add a virtual resistor in parallel with the capacitor. To eliminate the HF noise, a low-pass filter with time constant $T_v$ is also synthesized into the controller \cite{Meng2025}. A design example of the AD controller, denoted as $G_{ADv}(s)$, used in this paper is defined as
\begin{align}
\label{eq_GADv}
    G_{ADv}(s) = \frac{2.2\times10^{-6}s}{1/(4000\pi)s + 1}
\end{align}

Table \ref{tab_AD} shows that, with the designed $G_{ADv}(s)$, the system under droop control has HF OL poles located well within the LFP, far from the imaginary axis. Meanwhile, Fig. \ref{fig_NyquistAD}(c) shows that the Nyquist plot does not encircle the critical point (-1, 0). For a minimum phase system, it implies that the system is stable at high frequencies. In contrast, under droop-I control with the same $G_{ADv}(s)$, the system remains nonminimum phase, with two HF OL poles in the RHP. Therefore, although its Nyquist plot also does not encircle (-1, 0), the system is expected to suffer from HF instability. This indicates that a capacitor voltage-based AD design which effectively ensures HF stability of the system under droop control may be insufficient for the system under droop-I control if it cannot fully eliminate unstable OL poles. The result is consistent with the observations for inverter current- and grid current-based AD strategies.

The above analysis demonstrates that, compared to the droop control, although droop-I control enables the robust droop characteristics, it also introduces a stronger coupling between the RAP loop and the HF dynamics, resulting from the susceptibility to become nonminimum phase with unstable HF OL poles. To effectively damp HF resonances in an SL-GFM converter operating under droop-I control, this inherent HF OL unstable behavior must be explicitly considered in the design, which necessitates a higher level of damping.

\section{Comparative Study Cases}\label{sec_Case}

\begin{figure}[!t]
\centering
\includegraphics[width=\columnwidth]{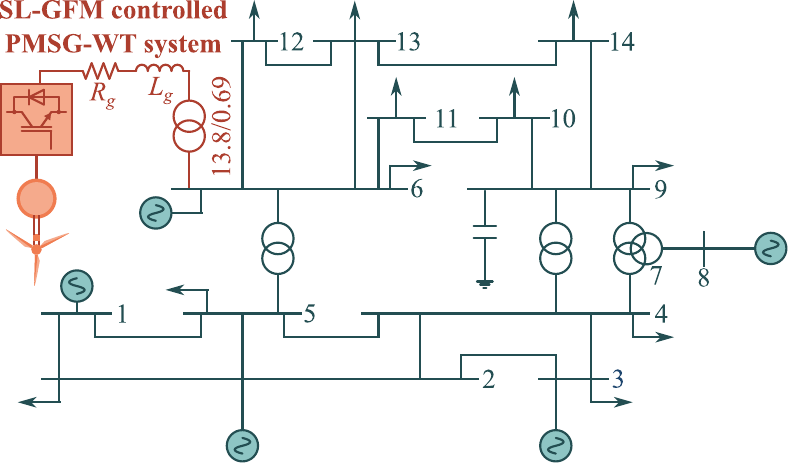}
\caption{IEEE 14-Bus Test System with connection of SL-GFM WT.}
\label{fig_IEEE14}
\end{figure}

This section compares the performance of SL-GFM WT under droop and droop-I controls when connected to the power system to highlight the potential HF oscillations, which are usually overlooked. Fig. \ref{fig_IEEE14} shows the IEEE 14-Bus Test System used to investigate the performance. A SL-GFM controlled PMSG-WT system is connected to Bus 6. The system parameters are the same as those listed in Table I unless given explicitly. 

First, the SL-GFM performance under droop and droop-I control strategies is compared. Thereafter, different AD strategies will be included and compared.

\subsection{Comparison of SL-GFM Controls}

\begin{figure}[!t]
\centering
\includegraphics[width=\columnwidth]{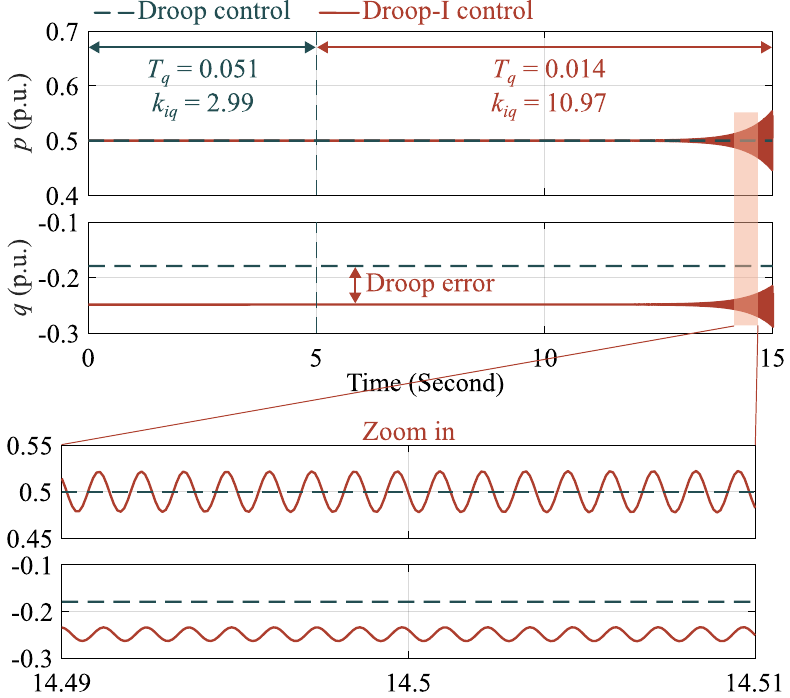}
\caption{Comparison of SL-GFM controls when RAP bandwidths increase with $L_g = 0.2$ p.u}
\label{fig_simulationSLBW}
\end{figure}

Fig. \ref{fig_simulationSLBW} shows the comparative results for changes in the RAP control bandwidth. Under droop control, the SL-GFM WT initially operates with $T_q=0.051$. At $t=5$ s, $T_{q}$ is decreased to 0.014 to increase the RAP control bandwidth. It is observed that the system remains stable throughout the entire process.

In comparison, under droop-I control, the SL-GFM WT initially operates stably with $k_{iq}=2.99$. At $t=5$ s, $k_{iq}$ is increased to 10.97 to increase the RAP control bandwidth. Following this change, the system starts to exhibit oscillations at approximately 830 Hz, indicating the emergence of HF resonance. As analyzed in Section \ref{Sec_comparison}, this HF instability originates from the transition of the system from an initial minimum phase condition to a nonminimum phase condition with OL unstable poles.

A droop error is also observed in Fig. \ref{fig_simulationSLBW}. Since the voltage at Bus 6 is higher than 1 p.u., the droop characteristics drive the WT system to absorb RAP. Due to the voltage drop across the filter, the system under droop control absorbs less RAP compared to the system under droop-I control.

The results indicate that droop-I control has strong coupling with the HF stability of the system. The systems are far more susceptible to HF resonance when the RAP control bandwidth is increased, compared to those under droop control.

\begin{figure}[!t]
\centering
\includegraphics[width=\columnwidth]{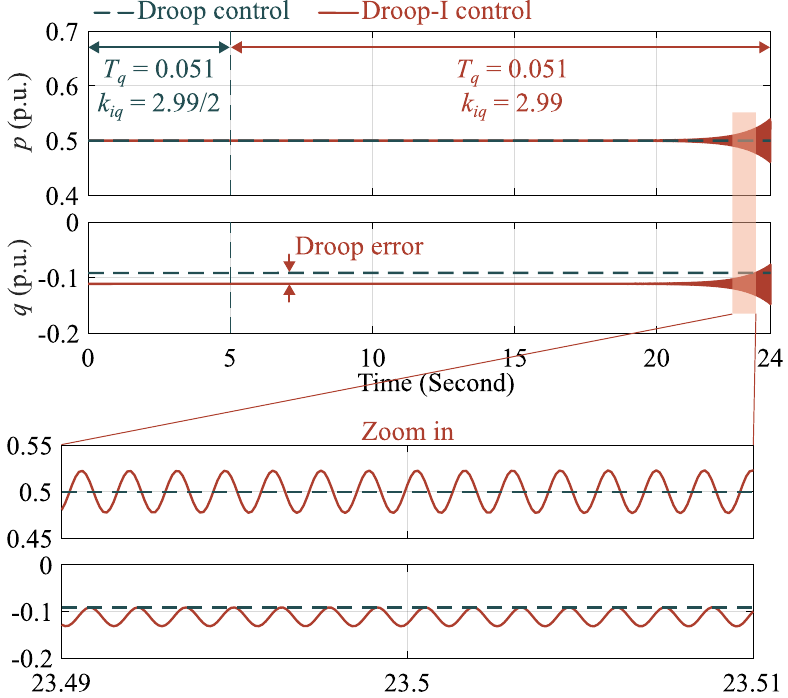}
\caption{Comparison of SL-GFM controls when RAP bandwidths increase with $L_g = 0.5$ p.u}
\label{fig_SimulationSLWG}
\end{figure}

Fig. \ref{fig_SimulationSLWG} compares the performances in a weaker grid with $L_g=0.5$ p.u. To enable stable startup under droop-I control, $k_{iq}$ is initially set to half of the original value. At $t=5$ s, $k_{iq}$ is restored to the original value of 2.99. However, the system then loses stability and oscillates at approximately 710 Hz, indicating the emergence of HF resonance. Likewise, as analyzed in Section \ref{sec_conclusion}, this HF instability arises from a transition of the system from an initial minimum phase condition to a nonminimum phase condition with OL unstable poles. Such transition is driven by the larger $L_g$ of 0.5 p.u., compared to 0.2 p.u. in Fig. \ref{fig_simulationSLBW}.

In comparison, the system under droop control remains stable with the original $T_q$ of 0.051 when $L_g=0.5$ p.u. A droop error is also observed in this case. Due to the larger $L_g$ , the capacitor voltage is lower than in the $L_g=0.2$ p.u. case. As a result, the SL-GFM WT system absorbs less RAP and the droop error is smaller as well.

The results reveal that HF stability of the systems under droop-I control is highly sensitive to grid strength. The system is with a much higher probability for HF resonance in weak grid conditions, compared to those systems under droop control.

\subsection{Comparison with Different AD Strategies}

In this section, a combination of larger RAP control bandwidth and weaker grid condition is considered, i.e., $T_q=0.014$, $k_{iq}=10.97$, and $L_g=0.5$ p.u. Based on the previous analysis and simulations, this configuration represents the worst-case scenario for HF stability in systems with droop-I control. The performance of various AD strategies is compared under both droop and droop-I control. To ensure stable startup, $T_q$ and $k_{iq}$ are initially set to 0.051 and 2.99, respectively. 

\subsubsection{Inverter Current-Based AD Strategy} 
\begin{figure}[!t]
\centering
\includegraphics[width=\columnwidth]{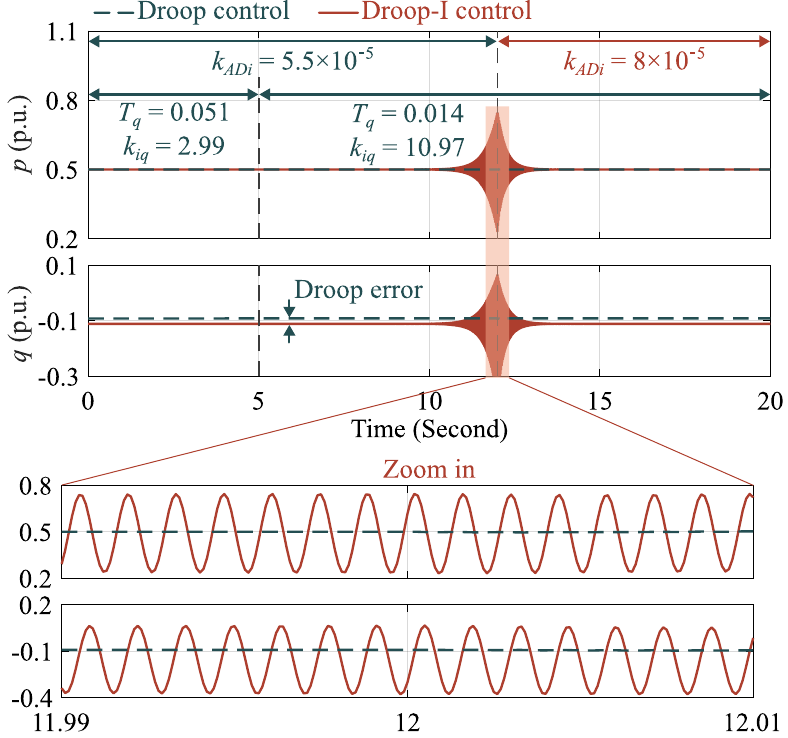}
\caption{Comparison of droop control and droop-I control with inverter current-based AD strategy.}
\label{fig_SimulationifAD}
\end{figure}

Fig. \ref{fig_SimulationifAD} compares the performance of the system under droop and droop-I control with inverter current-based AD strategy. The design example is the same as (\ref{eq_GADi}). After a stable startup, at $t=5$ s, $T_q$ and $k_{iq}$ are changed to $T_q=0.014$ and $k_{iq}=10.97$, corresponding to the tested worst-case condition. Following this change, the system under droop-I control exhibits oscillations at approximately 710 Hz, indicating the onset of HF resonance. In contrast, the system under droop control remains stable. Subsequently, at $t=12$ s, the gain of the AD controller, $k_{ADi}$, is increased by 45\%, from $5.5\times10^{-5}$ to $8\times10^{-5}$. This adjustment suppresses the HF resonance under droop-I control within 2 s.

\subsubsection{Grid Current-Based AD Strategy} 
\begin{figure}[!t]
\centering
\includegraphics[width=\columnwidth]{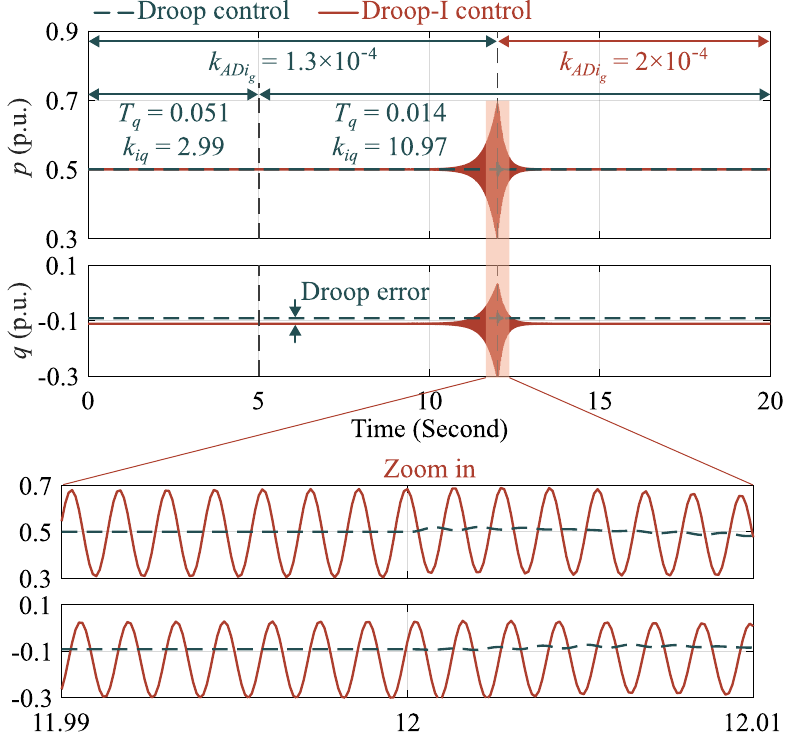}
\caption{Comparison of droop control and droop-I control with grid current-based AD strategy.}
\label{fig_SimulationiAD}
\end{figure}

Fig. \ref{fig_SimulationiAD} compares the performance of the system under droop and droop-I control with grid current-based AD strategy. The design example is the same as (\ref{eq_GADig}). As in the previous case, the system starts up stably. At $t=5$ s, $T_q$ and $k_{iq}$ are changed to $T_q=0.014$ and $k_{iq}=10.97$. Following this change, the system under droop-I control again exhibits HF resonance at approximately 710 Hz, whereas the system under droop control still remains stable. Thereafter, at $t=12$ s, the gain of the AD controller, $k_{ADi_g}$, is increased by 54\% from $1.3\times10^{-4}$ to $2\times10^{-4}$, and the HF resonance under droop-I control is suppressed within 2s.

\subsubsection{Capacitor Voltage-Based AD Strategy} 
\begin{figure}[!t]
\centering
\includegraphics[width=\columnwidth]{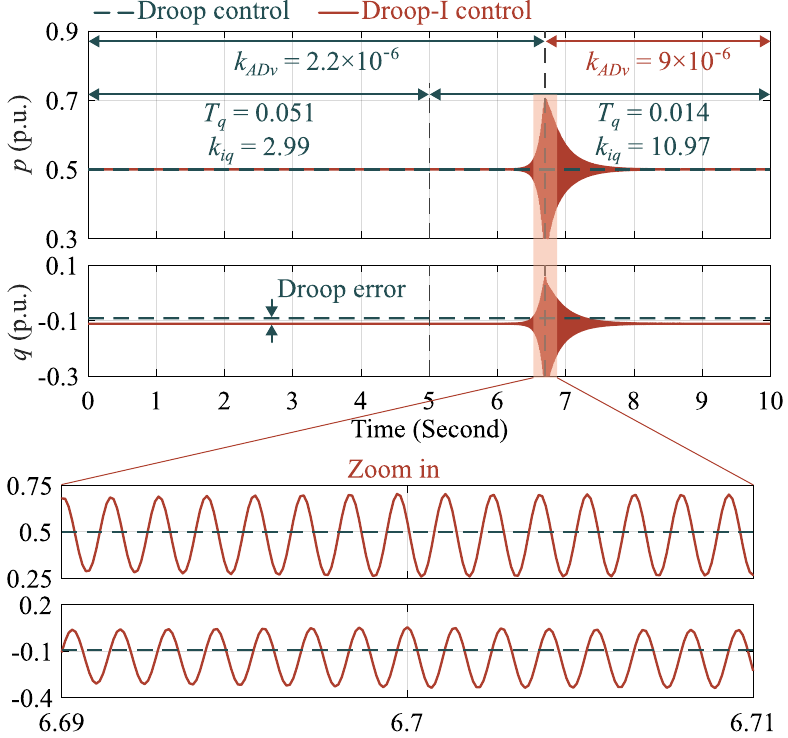}
\caption{Comparison of droop control and droop-I control with capacitor voltage-based AD.}
\label{fig_SimulationvAD}
\end{figure}

Fig. \ref{fig_SimulationvAD} compares the performance of the system under droop and droop-I control with capacitor voltage-based AD strategy. The design example is the same as (\ref{eq_GADv}). The same condition change is applied that, at $t=5$ s, $T_q$ and $k_{iq}$ are changed to $T_q=0.014$ and $k_{iq}=10.97$ from their original values 0.051 and 2.99, respectively. This change triggers HF resonance at approximately 710 Hz for the system under droop-I control, while the system under droop control still remains stable. Thereafter, at $t=6.7$ s, the gain of the AD controller, $k_{ADv}$, is increased by 300\% from $2.2\times10^{-6}$ to $9\times10^{-6}$, and the HF resonance under droop-I control is suppressed within 2 s.

Across all three AD strategies, systems under droop control maintain stability. In contrast, systems under droop-I control consistently suffers HF resonance, which can be mitigated by changing the designed controller gain. This confirms that an AD design which effectively ensures HF stability of the system under droop control may be insufficient for the system under droop-I control. As analyzed in Section \ref{sec_comparison_AD}, the resonance arises because the unstable OL poles are not fully eliminated. Therefore, to effectively damp HF resonances in an SL-GFM converter operating under droop-I control, this inherent HF OL unstable behavior must be explicitly addressed in the design, which necessitates a higher level of damping. Otherwise, the integration of the WTs will introduce new instability to the power systems.

\section{Conclusion}\label{sec_conclusion}
Renewable energy sources such as WTs are increasingly expected to provide GFM ability. This paper studies new instability phenomena introduced by the SL-GFM control strategies. Upon comparing the two widely used RAP controls, it appears that although both of them yield the droop behavior, they have substantial impact on HF stability of the power systems. The key findings are summarized as follows:
\begin{enumerate}
    \item While systems under droop control always remain minimum phase, the additional voltage magnitude feedback in droop-I control causes the systems to be always nonminimum phase with unstable OL HF poles in ideal inductive grid. This OL instability is a fundamental structural property of the system and cannot be mitigated by changing control or system parameters.
    \item In practical system with parasitic resistances, the system under droop-I control could be minimum phase. Nevertheless, it remains highly sensitive to parameter variations and is prone to transition to nonminimum phase. Notably, droop-I control exhibits strong coupling with the HF dynamics, where increasing the RAP loop bandwidth tends to drive the system into a nonminimum phase condition. Furthermore, a weaker grid similarly promotes this transition, thereby undermining one of the key advantages of GFM control, the ability to maintain stable operation under weak grid conditions.
    \item Compared to those under droop control, the OL unstable behavior causes systems under droop-I control to exhibit significantly weaker HF stability, as the HF resonances can still arise even when Bode's crossover criterion is satisfied.
    \item AD designs that effectively ensure HF stability in systems under droop control may be insufficient for systems under droop-I control if they do not fully eliminate unstable OL poles. This inherent HF OL unstable behavior must be explicitly addressed in the design with droop-I control, which necessitates a higher level of damping.
\end{enumerate}


%





\ifCLASSOPTIONcaptionsoff
  \newpage
\fi



\bibliographystyle{IEEEtran}
\bibliography{bibtex/bib/IEEEabrv,bibtex/bib/bibliography}
%

%








\end{document}